\documentstyle[11pt,epsf]{article}
\input epsf.tex
 1 
scaled\magstep 1  1

\def\lsim{\mathrel{\raise.3ex\hbox{$<$\kern-.75em\lower1ex\hbox{$\sim$}}}}

\def\gsim{\mathrel{\raise.3ex\hbox{$>$\kern-.75em\lower1ex\hbox{$\sim$}}}}

\begin{document}
\title{Coherent State for a Relativistic Spinless Particle}
\author{M. Haghighat$^{1,2}$\thanks{email: mansour@cc.iut.ac.ir}\ \ \
and \ \ A. Dadkhah$^{1}$\thanks{email:
adadkhah@cc.iut.ac.ir}\\ \\
  {\it $^1$Department of  Physics, Isfahan University of Technology (IUT)}\\
{\it Isfahan,  Iran,} \\{\it and}\\
  {\it $^2$Institute for Studies in Theoretical Physics and Mathematics (IPM)}\\
{\it P. O. Box: 19395-5531, Tehran, Iran.}}
\date{}
\maketitle

\begin{abstract}
The Klein-Gordon equation with scalar potential is considered. In
the Feshbach-Villars representation the annihilation operator for
a linear potential is defined and its eigenstates are obtained.
Although the energy levels in this case are not equally-spaced,
depending on the eigenvalues of the annihilation operator, the
states are nearly coherent and squeezed. The relativistic
Poschl-Teller potential is introduced.  It is shown that its
energy levels are equally-spaced. The coherence of time evolution
of the eigenstates of the annihilation operator for this
potential is evaluated.
\end{abstract}
\null
\section{Introduction}
Nowadays, the coherent state has found widespread application in
many branches of physics such as non linear optics, laser, nuclear
and particle physics. In the non-relativistic limit, the coherent
states are well known and one can produce them using three
different ways \cite{glaub}:
\begin{itemize}
\item[I.] {\it They can be constructed as eigenstates of the annihilation
operator.}
\item[II.]{\it They can be defined as quantum states with minimum uncertainty
relationship.}
\item[III.] {\it They can be obtained by operating the Glauber's displacement
operator on the vacuum state.}
\end{itemize}
  The above definitions are known to be equivalent for the Schr\"{o}dinger
equation with the harmonic-oscillator potential. In references
[2-12] coherent states for systems other than the
harmonic-oscillator are also investigated.

To extend the {\it coherence} to relativistic region one may use
manifestly covariant equations \cite{ald}. Of course in this way,
it is difficult to describe the wave functions in terms of four
vectors.  Thus, in many areas of physics, one may use the
Schr\"{o}dinger equation with relativistic corrections or the
Klein-Gordon (KG) or Dirac equation with specific potentials to
describe a relativistic particle or even bound state in a
non-covariance manner.  To describe the relativistic coherent
states using the KG or Dirac equation one encounters particle pair
creation that makes one particle interpretation nonsense.  To
avoid this problem one should consider the theory under
conditions that the pair creation is impossible.  Furthermore in
relativistic quantum mechanics the consistency of the one
particle description is limited and in this case the only valid
operators are \textit{even} operators which do not mix different
charge states.  In this direction authors of reference \cite{klu}
have studied a spin 0 charged particle in two cases: free
particle and a particle in a constant homogeneous magnetic field.

 In this
paper we examine the possibility of coherence for KG equation
with a pure scalar potential and adopt the definition $I$ to extend
the coherence to the relativistic region.  Such a potential describes a position dependent
valence and conduction-band edge of semiconductors near the $\Gamma$ or $L$ point
in the Brillouin zone.  In fact the position dependent mass characterizes a position
dependent band gap.  Examples for such materials are given in ref. \cite{Doni}.  \\
In section 2 we obtain $x$ and $p$ operators in the
Feshbach-Villars representation for a general scalar potential.
Subsequently we derive eigensolutions of a linear scalar
potential in the Schr\"{o}dinger representation to show that pair
creation does not occur.  Consequently we construct the
eigenstates of the annihilation operator for this potential. The
time evolution of the states is numerically discussed in section
3. In section 4 we introduce a potential with equally spaced
energy levels and examine their coherence with time.  The
resolution of unity for the obtained states is also discussed in
this section.
\section{KG equation for a Scalar potential}
In order to explore the coherence of a relativistic particle in
the simplest case we consider the KG equation in one dimension
with a vector and a scalar potential ($\hbar=c=1$)[15,16]
\begin{equation}
\left[{\partial^2\over {\partial x^2}}+ (i{\partial\over
{\partial t}}-V(x))^2-(m+S(x))^2\right]\psi(x,t)=0,
\end{equation}
where $m$ is the mass of the particle, $S(x)$ is the scalar
potential and $V(x)$ is the time component of a 4-vector
potential.  In non-relativistic limit, Eq.(1) leads to a
Schr\"{o}dinger equation with an effective potential as $S+V$.  In
fact this equation gives all relativistic corrections to the
Schr\"{o}dinger equation with such an effective potential.  For a
pure scalar potential we have

\begin{equation}
H_s\psi(x,t)=-{1\over 2m}{\partial^2\over {\partial t^2}}
\psi(x,t), \label{kg1}
\end{equation}
in which
\begin{equation}
H_s=-{1\over {2m}}{d^2\over {dx^2}}+ {(m+S(x))^2\over 2m}.
\end{equation}
 Now by introducing two functions
 $\phi$ and $\chi$ and the ansatz
\begin{equation}
\psi=\varphi+\chi\hspace{1cm};\hspace{1cm}i{\partial\psi\over\partial
t}=m(\varphi-\chi),
\end{equation}
we transform Eq.(\ref{kg1}) to the Schr\"{o}dinger form as follows
\begin{equation}
i{\partial\Psi\over\partial
t}=\left[(\sigma_3+i\sigma_2)(H_s-{m\over
2})+\sigma_3m\right]\Psi, \label{schkg}
\end{equation}
where $\sigma_i^s$, $i=1,2,3$, are the Pauli matrices and $\Psi$
is a two components column vector as
\begin{equation}
\Psi=\left(\begin{array}{c}\varphi\\\chi\end{array}\right).
\end{equation}
It is easy to show that each component of $\Psi$ individually
satisfies the KG equation i.e. Eq.(\ref{kg1}).  If we define
$u_n$ as eigenstate of the Hamiltonian $H_s$ with corresponding
eigenvalue $\epsilon_n$ with the ansatz
\begin{equation}
\Psi=\left(\begin{array}{c}\varphi_0\\
\chi_0\end{array}\right)u_ne^{[-iEt]}, \label{ansa}
\end{equation}
one obtains the positive and negative energy solutions as follows
\begin{equation}
\Psi^\pm=N_\pm\left(\begin{array}{c}m\pm E_n\\
m\mp E_n\end{array}\right)u_n{e}^{[\mp iE_nt]}, \label{un}
\end{equation}
where $N_\pm$ are the appropriate normalization constants and
\begin{equation}
E_n=\sqrt{2m\epsilon_n}.
\end{equation}
 Now an operator
\begin{equation}
U={(m+{\cal H})-\sigma_1(m-{\cal H})\over\sqrt{4m{\cal H}}},
\end{equation}
in which
\begin{equation}
{\cal H}=\sqrt{2mH_s},
\end{equation}
 can transform the Schr\"{o}dinger representation to the
Feshbach-Villars representation therefore it can be shown that the
Hamiltonian in this representation is
\begin{equation}
H_{FV}=\sigma_3{\cal H}=\sigma_3\sqrt{2mH_s},
\end{equation}
which is even and the even parts of operators $x$ and $p$ are
also the same as the corresponding operators in the
non-relativistic region:
\begin{equation}
x_{FV}=iU{\partial\over\partial p}U^{-1}=i{\partial\over\partial
p}-i{p\sigma_1\over 2E_n^2},\label{xfv}
\end{equation}
for x-operator in momentum space and
\begin{equation}
p_{FV}=-iU{\partial\over\partial x}U^{-1}=-i{\partial\over\partial
x}+i{({\partial S\over \partial x})(m+S)\sigma_1\over
2E_n^2},\label{pfv}
\end{equation}
for p-operator in configuration space.  Now we are ready to study
the coherent states for potentials without pair creation.  First,
we choose a linear potential as follows
\begin{equation}
S(x)=k |x|- m ,
\end{equation}
where $k$ is a coupling constant.  Introducing parameters
$\epsilon_n =E^2/(2m)$ and $\omega=k/m$,  by using the ansatz of
Eq.(\ref{ansa}), Eq.(\ref{kg1}) can be written in a familiar form
-the Schr\"{o}dinger equation with a harmonic-oscillator
potential. Therefore one can easily find
\begin{equation}
\Psi_n^\pm\propto\left(\begin{array}{c}m\pm E_n\\
m\mp E_n\end{array}\right)e^{-{m\omega x^2\over 2}}H_n({\sqrt {mw}
}x ){e}^{[\mp iE_nt]},
\end{equation}
in which $H_n(x)$ is the Hermite polynomial of order $n$ and
\begin{equation}
E=\pm\sqrt{(2n+1)k}=\pm E_n , \label{hoe}
\end{equation}
where $n=0, 1, 2, ...$.  One can easily see from Eq.(\ref{hoe})
that pair creation in this case never occurs.   For a pure scalar
potential the KG equation is independent of the sign of $E$. In
fact the scalar interaction is independent of the charge of the
particle and has the same effect on particles and
anti-particles.  Thus, naively speaking, this kind of potential
can describe the confining part of the potential of a quarkonium.

Since the even part of an operator is a measurable quantity
therefore Eq.(\ref{xfv}) and Eq.(\ref{pfv}) allow us to consider
the usual annihilation operator for a non-relativistic harmonic
oscillator potential to construct the coherent state for our
linear potential. Therefore the coherent state $|\alpha\rangle$ at
t=0 in the Feshbach-Villars representation in terms of the energy
eigenstate for a non-relativistic harmonic oscillator $|n\rangle$,
can be obtained as
\begin{equation}
|\alpha,+\rangle=e^{-|\alpha|^2\over 2}\sum_n{\alpha^n\over
{\sqrt{n!}}}\left(\begin{array}{c}1\\0\end{array}\right)|n\rangle
,\label{alfa}
\end{equation}
for positive energy state and
\begin{equation}
|\alpha,-\rangle=e^{-|\alpha|^2\over 2}\sum_n{\alpha^n\over
{\sqrt{n!}}}\left(\begin{array}{c}0\\1\end{array}\right)|n\rangle
,
\end{equation}
for negative energy state. Thus the state given in
Eq.(\ref{alfa}), when we restrict ourselves to the positive
energy states, evolves as
\begin{equation}
|\alpha(t),+ \rangle=e^{-|\alpha|^2\over 2}\sum_n{\alpha^n\over
{\sqrt{n!}}}\left(\begin{array}{c}1\\0\end{array}\right)
e^{-iE_nt} |n\rangle ,\label{alfat}
\end{equation}
where $E_n=\sqrt {(2n+1)k}$.  Although $|\alpha\rangle$ is
coherent, its time evolution is not the eigenstate of the
annihilation operator.  Indeed, this is due to the
non-equally-spaced of the energy levels given in Eq.(\ref{hoe}).\\

One of the significant properties of the coherent states is an
over completeness relation or resolution of unity as follows
\begin{equation}
\int|\alpha\rangle d\mu (\alpha)\langle\alpha|=1 ,\label{resol}
\end{equation}
where the measure $ d\mu (\alpha)$ for the obtained states in
this section (i.e. Eq.(\ref{alfa})) can be easily determined as
\begin{equation}
d\mu (\alpha)={d[Re\alpha]d[Im \alpha]\over \pi} .\label{meas}
\end{equation}
\section{The Coherence of the Solutions for Linear Potential}
To explore the behavior of the solutions given in Eq.(\ref{alfat})
we first, examine the expectation values of the appropriate
quantities as follows
$$\langle x\rangle=\sqrt{2\over k}e^{-|\alpha|^2}\sum_n{
\left[a\cos(E_n-E_{n+1})t - b\sin(E_n-E_{n+1})t\right]  }
{|\alpha|^{2n}\over n!},$$
$$\langle p\rangle=\sqrt{2 k}e^{-|\alpha|^2}\sum_n[a\sin(E_n-E_{n+1})t+b\cos(E_n-E_{n+1})t]
 {|\alpha|^{2n}\over
n!},$$
$$\langle x^2\rangle={1\over k}(|\alpha|^2+{1\over2})$$
$$+{e^{-|\alpha|^2}\over
k}\sum_n[(a^2-b^2)\cos(E_n-E_{n+2})t-2ab\sin(E_n-E_{n+2})t]
{|\alpha|^{2n}\over n!},$$
$$\langle p^2\rangle=k(|\alpha|^2+{1\over2})$$
\begin{equation}
-ke^{-|\alpha|^2}\sum_n[(a^2-b^2)\cos(E_n-E_{n+2})t-2ab\sin(E_n-E_{n+2})t]
{|\alpha|^{2n}\over n!},
\end{equation}
where
\begin{equation}
\alpha=a+ib.\label{eigen}
\end{equation}
Time variations of $\Delta x$, $\Delta p$ and $\Delta x*\Delta
p$, for various values of $\alpha$, are illustrated in
Figs.(1)-(11).\\
\begin{figure}
\centerline{\epsfxsize=3in\epsffile{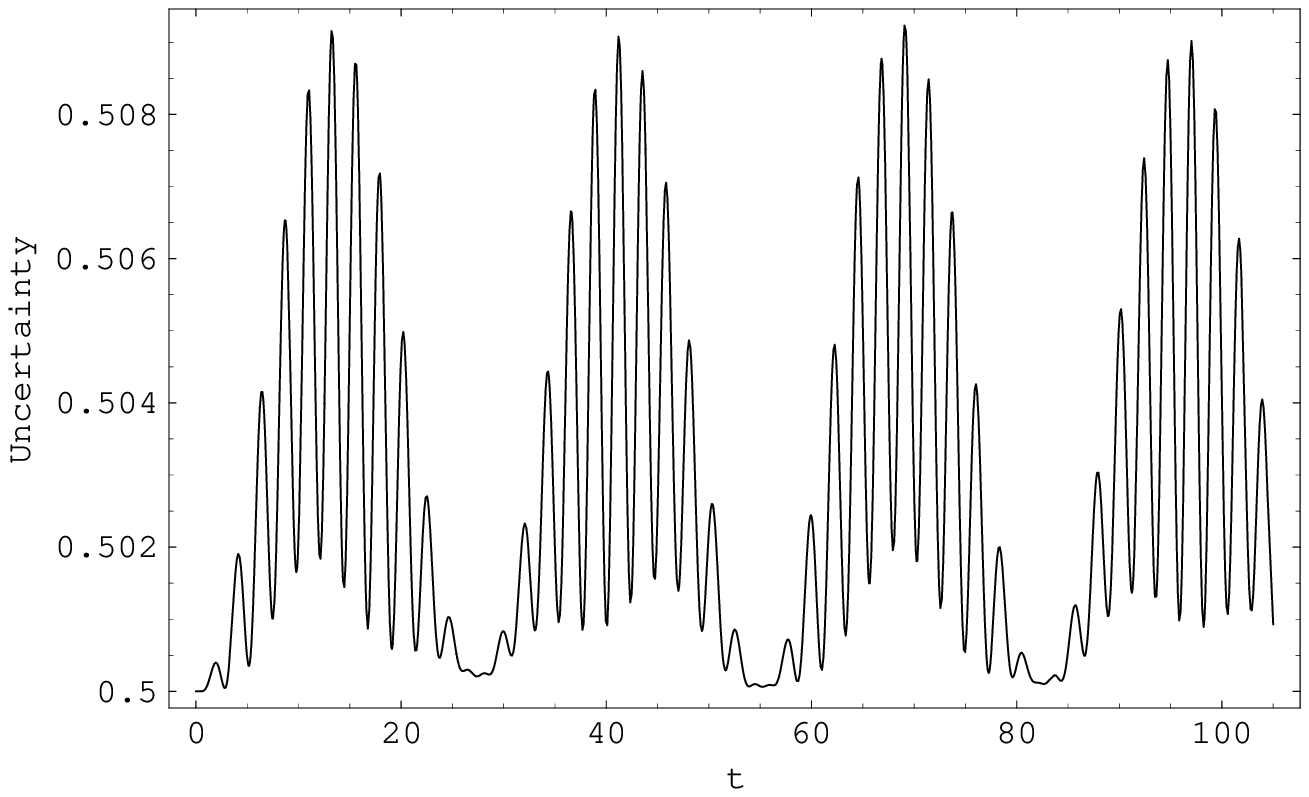}} \caption{$\Delta
x*\Delta p$ for $\alpha=0.1+0.2i$ as a function of time. }
\end{figure}
\begin{figure}
\centerline{\epsfxsize=3in\epsffile{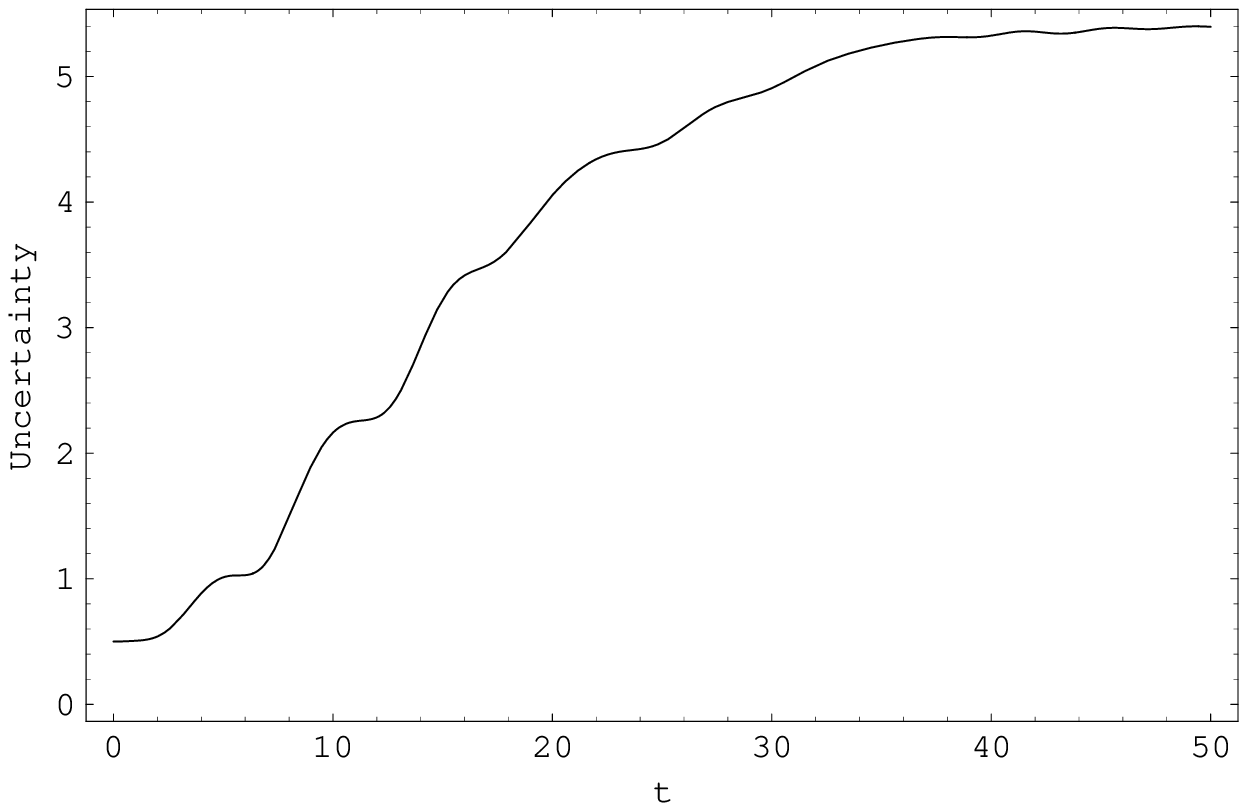}}
\caption{$\Delta x*\Delta p$ for $\alpha=1+2i$ as a function of
time for $0< t <50$. }
\end{figure}
\begin{figure}
\centerline{\epsfxsize=3in\epsffile{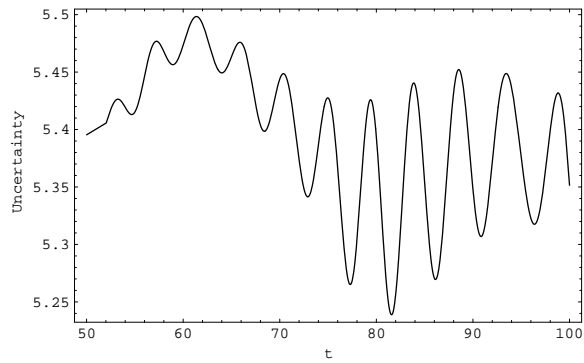}}
\caption{$\Delta x*\Delta p$ for $\alpha=1+2i$ as a function of
time for $50< t <100$. }
\end{figure}
For $\alpha=\alpha_1=0.1+0.2i$ the value of $\Delta x*\Delta p$
oscillates between 0.50 and 0.51 and the state is nearly
coherent, Fig.(1).  In contrast $\Delta x*\Delta p$ for
$\alpha=\alpha_2=1+2i$ increases with time though its value for
long time is bounded and oscillates about a certain value,
Figs.(2-3).\\
\begin{figure}
\centerline{\epsfxsize=3in\epsffile{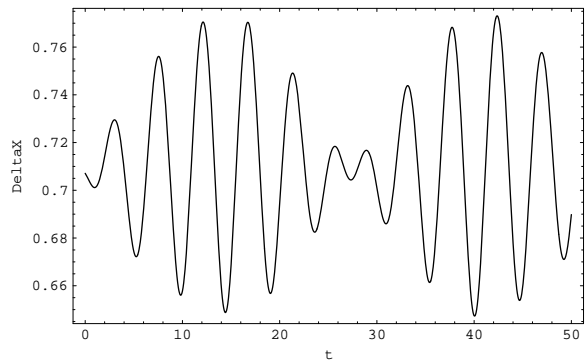}} \caption{
$\Delta x$ for $\alpha=0.1+0.2i$ as a function of time. }
\label{a6a}
\end{figure}
\begin{figure}
\centerline{\epsfxsize=3in\epsffile{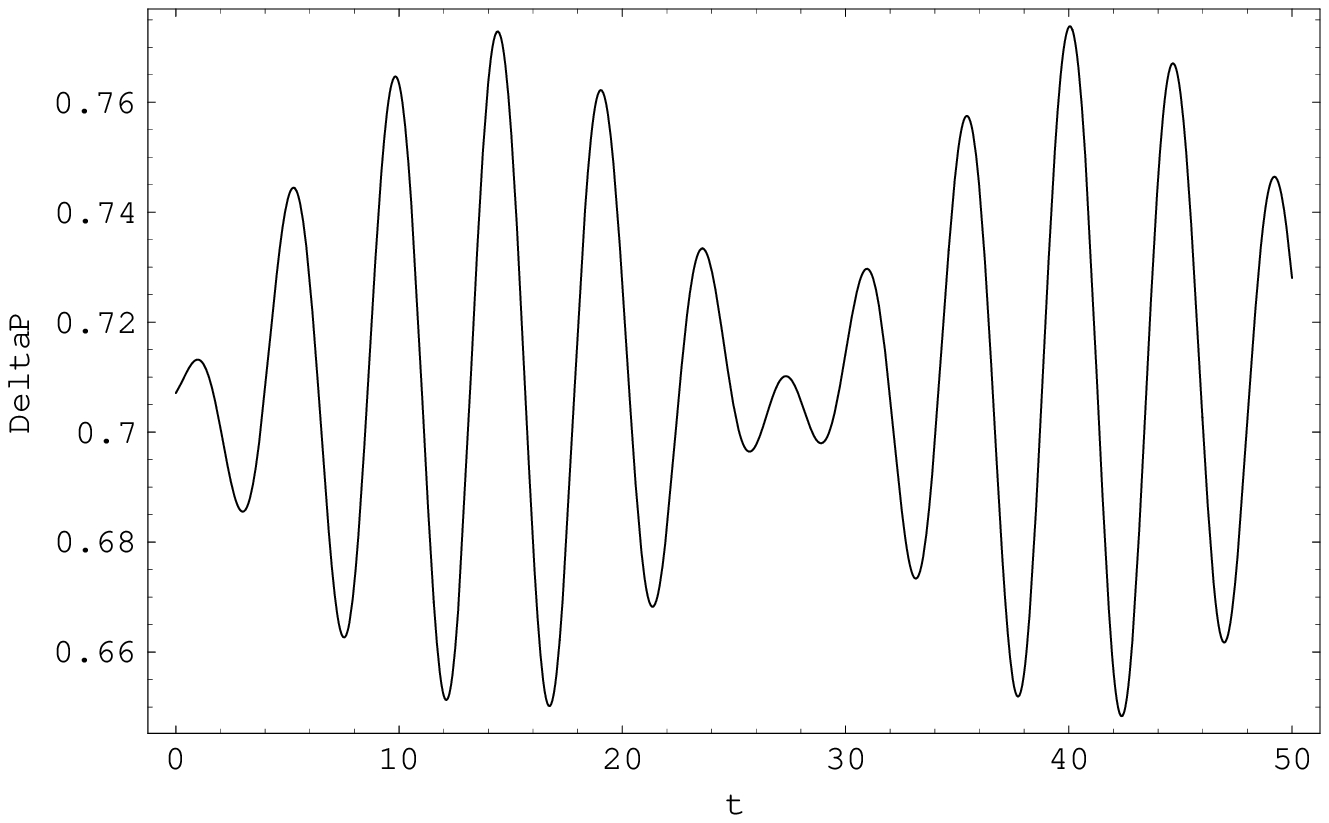}} \caption{
$\Delta p$ for $\alpha=0.1+0.2i$ as a function of time. }
\end{figure}
\begin{figure}
\centerline{\epsfxsize=3in\epsffile{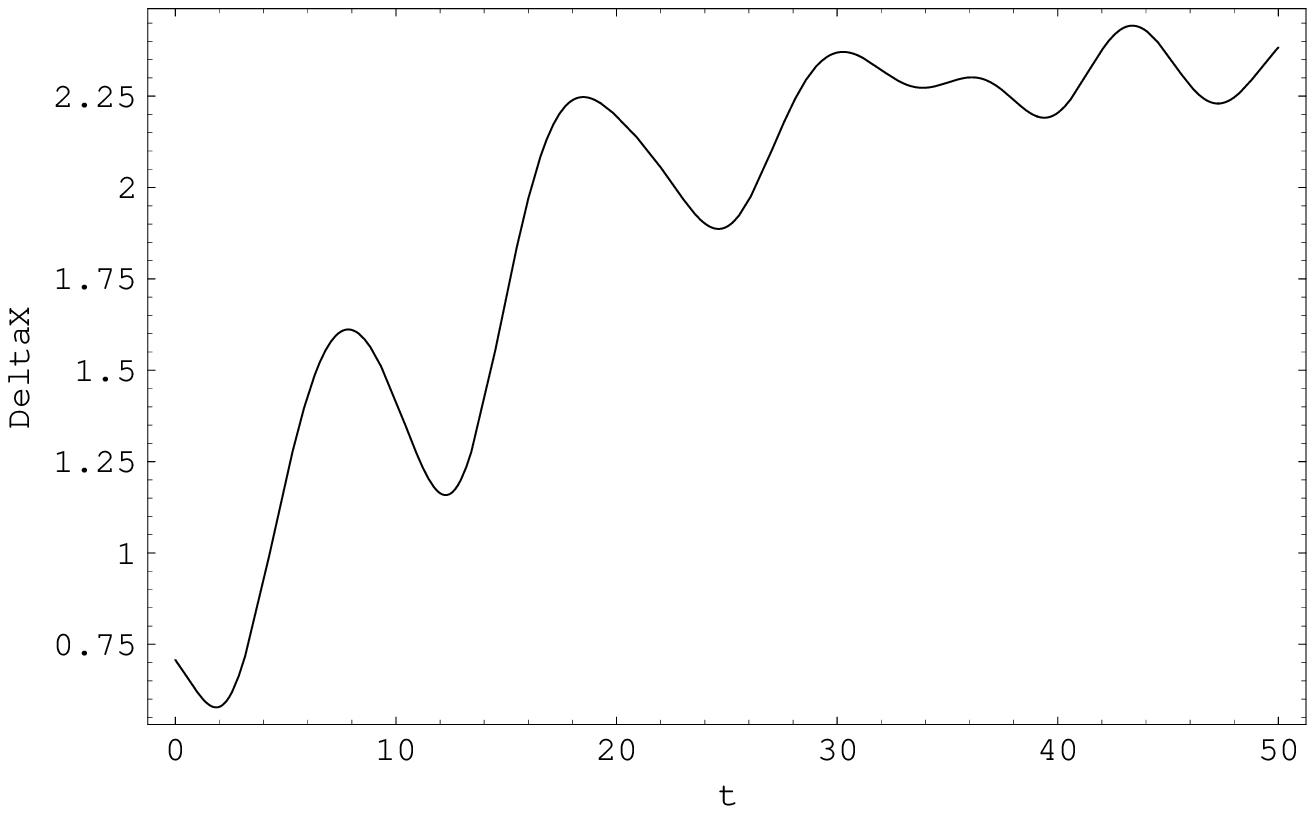}} \caption{
$\Delta x$ for $\alpha=1+2i$ as a function of time. }
\end{figure}
\begin{figure}
\centerline{\epsfxsize=3in\epsffile{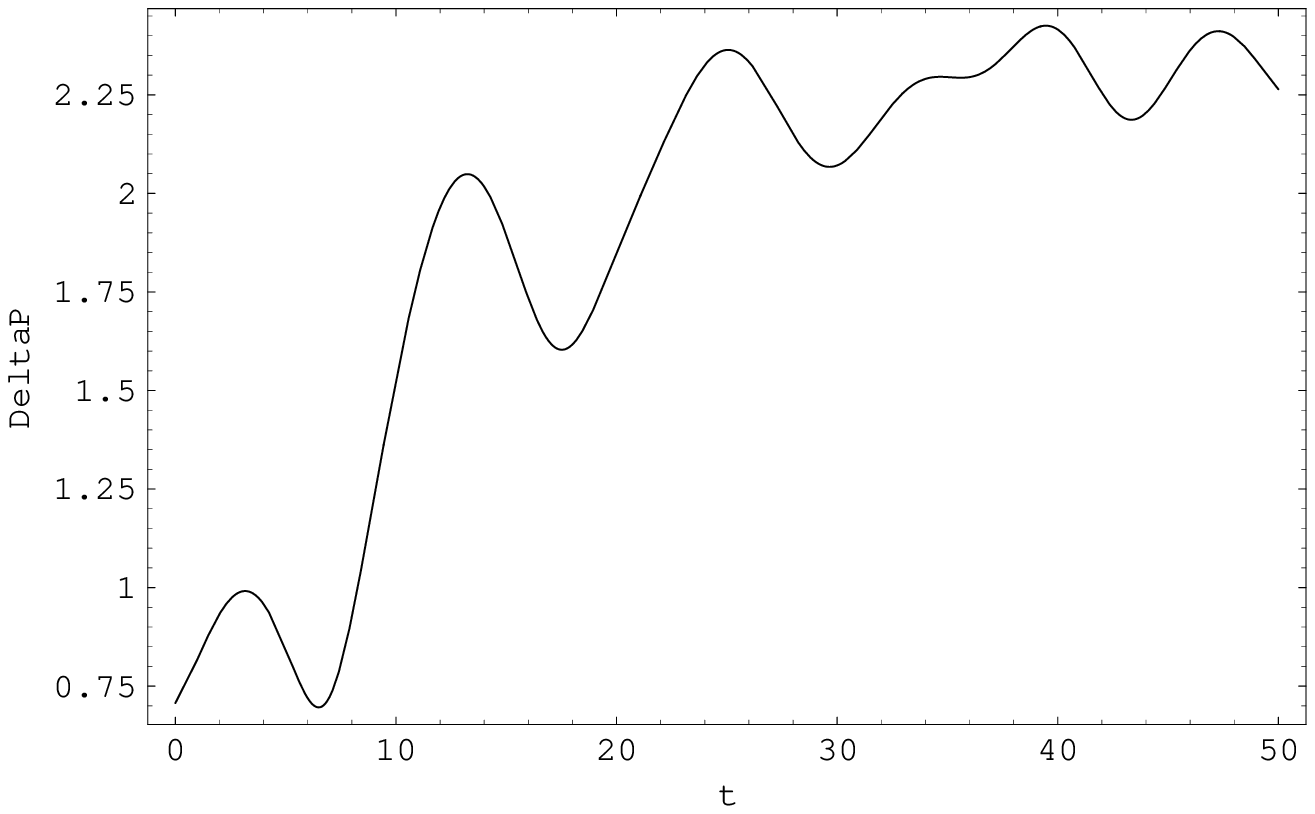}}
\caption{$\Delta p$ for $\alpha=1+2i$ as a function of time. }
\end{figure}
The numerical values of $\Delta x$ and $\Delta p$ in both cases
are not constant as shown in Figs.(4-7).  When $\alpha=\alpha_1$
they oscillate almost periodically about their minimum values.
Furthermore, their oscillations are exactly out of phase and the
states are squeezed, Figs.(4-5).  When $\alpha$ increases to
$\alpha_2$, the values of $\Delta x$ and $\Delta p$ increase with
time and finally oscillate about a finite value, Figs.(6-7).\\

The expectation values of $x$ and $p$ for $\alpha=\alpha_1$ and
$\alpha=\alpha_2$ are plotted in Figs.(8-9) and Figs.(10-11),
respectively.  In the case $\alpha=\alpha_1$ the values $\langle
x\rangle$ and $\langle p\rangle$ oscillate with an amplitude and
a period nearly constant which is very similar to a classical
harmonic oscillator.  Again for $\alpha=\alpha_2$ the situation
is exactly different and the values of $\langle x\rangle$ and
$\langle p\rangle$ oscillate with a variable period and amplitude.  \\
In the evaluation of the various quantities we have summed terms
up to $n=50$ and assumed $k=1$, though the general behaviors do
not depend on $k$.

\begin{figure}
\centerline{\epsfxsize=3in\epsffile{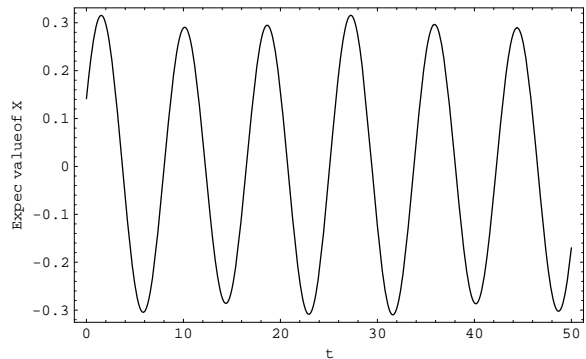}} \caption{The
expectation value of x for $\alpha=0.1+0.2i$ as a function of
time. }
\end{figure}
\begin{figure}
\centerline{\epsfxsize=3in\epsffile{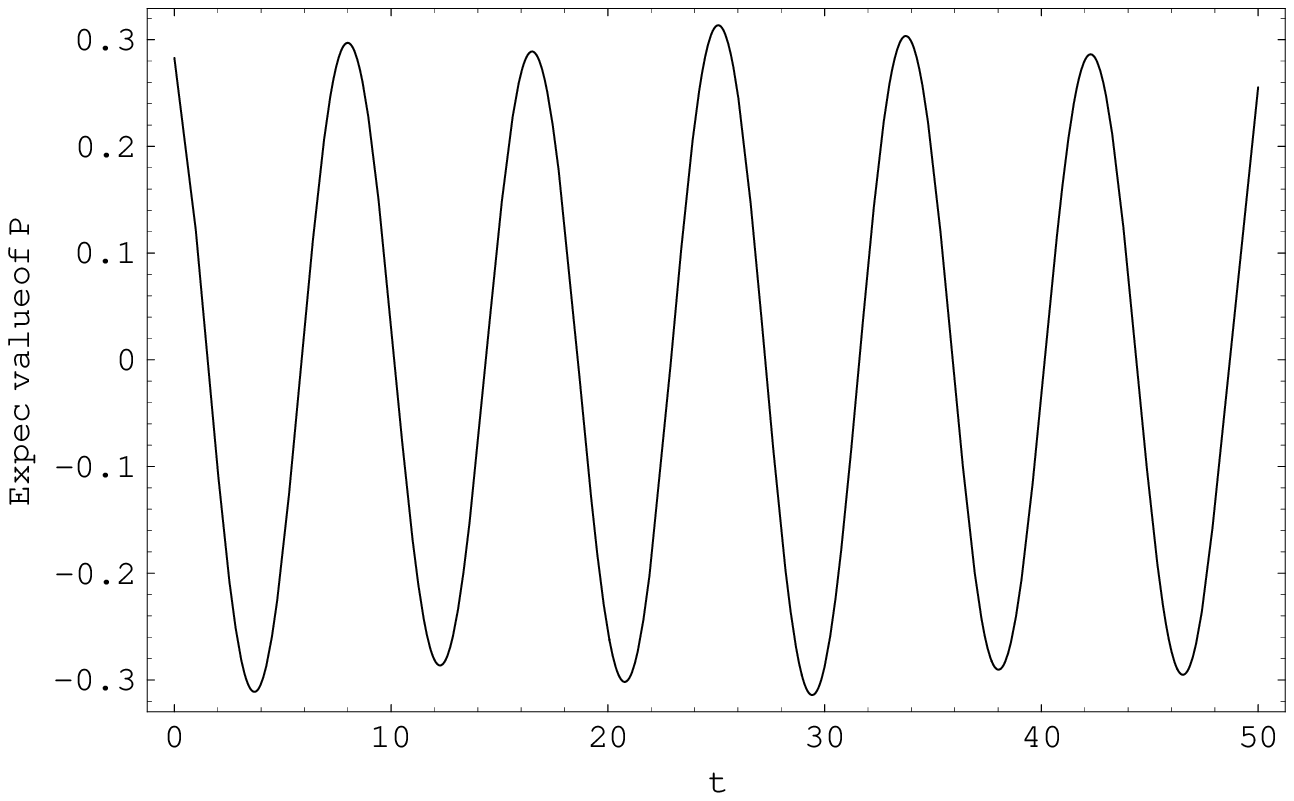}} \caption{The
expectation value of p for $\alpha=0.1+0.2i$ as a function of
time. }
\end{figure}
\begin{figure}
\centerline{\epsfxsize=3in\epsffile{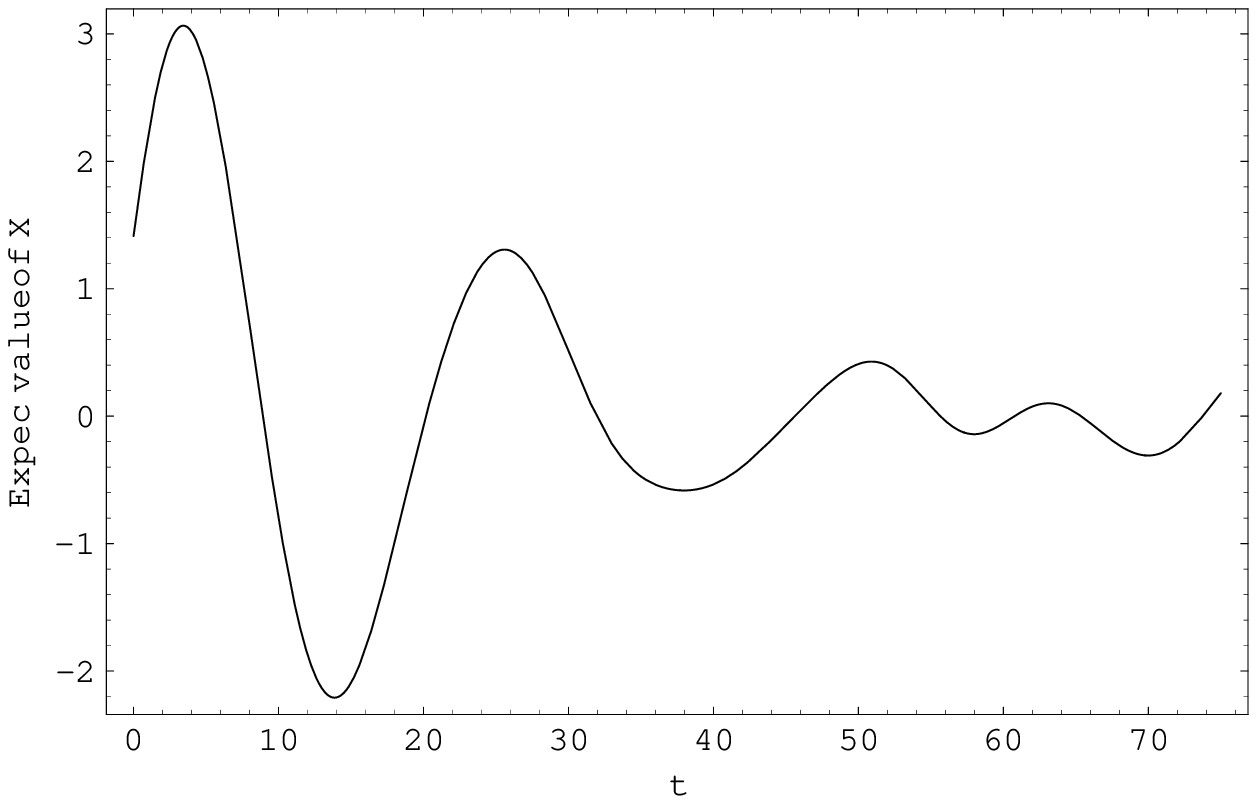}} \caption{The
expectation value of x for $\alpha=1+2i$ as a function of time. }
\end{figure}
\begin{figure}
\centerline{\epsfxsize=3in\epsffile{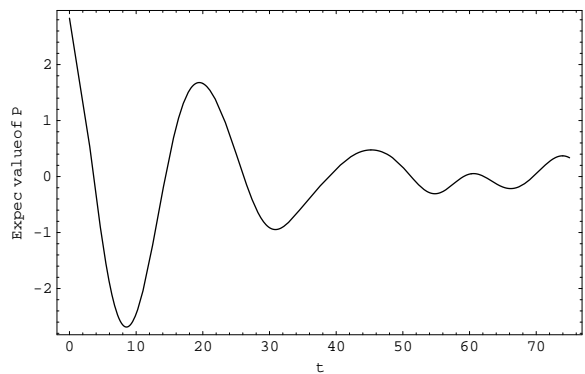}} \caption{The
expectation value of p for $\alpha=1+2i$ as a function of time. }
\end{figure}
\newpage
\section{Potentials with Equally-Spaced Levels }
In the preceding section we saw that a relativistic scalar
particle with a linear scalar potential, in general, is not
coherent. The main reason for deviation from coherence is the fact
that the eigenenergies are not equally spaced.  Now, we would like
to construct potentials with equally-spaced energy levels. For
this purpose we rewrite $H_s$ as follows
\begin{equation}
H_s=-{1\over {2m}}{d^2\over {dx^2}}+ {U^2+m^2\over
2m},\label{kgpt}
\end{equation}
where $U^2(x)=S^2(x)+2mS(x)$.  We now introduce a potential as
follows
\begin{equation}
U(x)=m\tan({\omega x}),\label{pt}
\end{equation}
in the region ${-\pi\over 2\omega}\le x\le{\pi\over 2\omega}$ and
is infinite elsewhere, or in the same region
\begin{equation}
S(x)=-m \pm {m\over {\cos({\omega x}})},\label{pt1}
\end{equation}
where $\omega$ is a constant. Substituting this potential in
Eq.(\ref{kgpt}) and using ansatz (\ref{ansa}) equation (\ref{kg1})
leads to the non-relativistic Poschl-Teller (PT) equation
\cite{pos} and consequently $u_n$ in Eq.(\ref{un}) can be
obtained as
\begin{equation}
u_n(x)=\left[{{\omega (\lambda +n)\Gamma (2\lambda +n)}\over
{\Gamma (n+1)}}\right]^{1\over 2} (\cos\omega x)^{1\over
2}P_{n+\lambda -{1\over 2}}^{{1\over 2}-\lambda}(\sin\omega x)
\end{equation}
and for the energy eigenvalues one has
\begin{equation}
E=\pm E_n=\pm {\omega}(n+\lambda)\:\:\:\:;\:\:\:n=0,1,2,...
 \label{pten}
\end{equation}
where
\begin{equation}
\lambda={1\over 2}+{1\over 2}\sqrt{{4m^2\over \omega^2} +1}.
\end{equation}
It should be noted that although in the relativistic PT
potential, Eq.(\ref{pt1}), the energy levels are equally-spaced
but the PT potential in the Schr\"{o}dinger equation has
non-equally-spaced levels [3,4]. Here again pair creation does
not occur and one-particle sector of the theory is applicable.
Now we need the annihilation operators to construct the coherent
states.  To this end, one can use the Schr\"{o}dinger method to
determine the ladder operators as follows \cite{niesi}:
\begin{equation}
A_{\pm}=(\sin \omega x)\left[{2m(H+{m\over 2})\over
{\omega}}\right]^{1\over2}\mp {1\over \omega}(\cos\omega
x){d\over dx}
\end{equation}
where $H$ is the Hamiltonian of the Schr\"{o}dinger equation for
the PT potential.  Therefore in the Feshbach-Villars
representation we have
\begin{equation}
A_{\pm}{\psi}^{FV}_n(t=0)=(n+\lambda)D(n-{1\over 2}{\pm}{1\over
2},\lambda){\psi}^{FV}_{n\pm 1}(t=0)
\end{equation}
where ${\psi}^{FV}$ is the wave function in the Feshbach-Villars
representation
\begin{equation}
\psi^{FV}_n(t)=N_\pm\left(\begin{array}{c}1\pm 1\\
1\mp 1\end{array}\right)u_n{e}^{[\mp iE_nt]},
\end{equation}
in which $\pm$ stands for positive and negative energy solutions,
respectively, and
\begin{equation}
D(n,\lambda)=\left[{{(n+1)(2\lambda+n)}\over
{(n+\lambda)(n+1+\lambda)}}\right]^{1\over 2}.
\end{equation}
Let us now determine ${\psi}_{\alpha}$ as the eigenstate of the
annihilation operator $A_-$, with eigenvalue $\alpha$, by using
an expansion on the states ${\psi}^{FV}_n$:
\begin{equation}
{\psi}_\alpha=\sum_nc_n{\psi}^{FV}_n(t=0),
\end{equation}
we then have
\begin{equation}
A_-{\psi}_\alpha=\sum_nc_n(n+\lambda)D(n-1,\lambda){\psi}^{FV}_{n-1}(t=0)=
\alpha{\psi}_\alpha ,
\end{equation}
Therefore
\begin{equation}
c_{n+1}=\alpha\left[{{(n+\lambda)}\over
{(n+1)(2\lambda+n)(n+1+\lambda)}}\right]^{1\over 2}c_n,
\end{equation}
and, finally ${\psi}_\alpha$ can be obtained as
\begin{equation}
{\psi}_\alpha=N_\alpha\sum_n{\alpha}^n\left[{{\lambda\Gamma(2\lambda)}
\over {n!(n+\lambda)\Gamma(2\lambda+n)}}\right]^{1\over
2}{\psi}^{FV}_n(t=0),\label{alpt}
\end{equation}
where
\begin{equation}
N_\alpha=[\lambda\Gamma(2\lambda)S(\alpha)]^{-{1\over 2}},
\end{equation}
and
\begin{equation}
S(\alpha)=\sum_n{{|\alpha|^{2n}}\over
{n!(n+\lambda)\Gamma(2\lambda+n)}}.
\end{equation}
Now we examine the evolution of ${\psi}_\alpha$ at every
instant.  For this purpose we restrict ourselves to the positive
energy states which implies
\begin{equation}
{\psi}_\alpha(t)=N_\alpha\sum_n{\alpha}^n\left[{{\lambda\Gamma(2\lambda)}
\over {n!(n+\lambda)\Gamma(2\lambda+n)}}\right]^{1\over
2}e^{-iE_nt}{\psi}^{FV}_n(t=0),\label{alptt}
\end{equation}
where $E_n$ is given in Eq.(\ref{pten}).  In this case since the
eigenenergies are equally spaced one has
\begin{equation}
{\psi}_\alpha(t)=e^{-i\omega\lambda t}{\psi}_{\alpha(t)},
\end{equation}
where
\begin{equation}
\alpha(t)=e^{-i\omega t}{\alpha}.
\end{equation}
Therefore, the coherent state wave packet remains an eigenvector
of $A_-$ with an eigenvalue $\alpha e^{-i\omega t}$.

Now we investigate the resolution of unity for the obtained
states.  As (\ref{alpt}) shows the measure in this case is
different from the standard form of (\ref{meas}) but one can
follow the method of references \cite{klu} and \cite{sudar} to
find
\begin{equation}
d\mu(\alpha)=S(|\alpha|^2){\cal W}(|\alpha|^2){d[Re\alpha]d[Im
\alpha]\over \pi},\label{measpt}
\end{equation}
where ${\cal W}(|\alpha|^2)$ can be determined by solving the
following equation
\begin{equation}
\int_0^{\infty}|\alpha|^{2n}{\cal
W}(|\alpha|^2){d|\alpha|^2}={n!(n+\lambda)\Gamma(2\lambda+n)}.\label{measpt1}
\end{equation}
\section*{summary}

To summarize, in the Feshbach-Villars representation we obtained
$x$ and $p$ operators for a general scalar potential.  The even
parts of these operators coincide with their counterparts in the
Schr\"{o}dinger representation, see Eqs.(\ref{xfv},\ref{pfv}).
Consequently, we constructed the annihilation operator coherent
state for a purely linear scalar potential for a relativistic
spinless particle.  The properties of these states in $(1+1)$
dimension are:
\begin{itemize}
\item[i.] The eigenfunctions for the small eigenvalues Eq.(\ref{eigen}) are
quasi-coherent and squeezed. Furthermore $\langle x\rangle$ and
$\langle p\rangle$ oscillate like a classical harmonic
oscillator. In the other words for the small values of $\alpha$,
which is not the non-relativistic limit of Eq.(\ref{kg1}), the
states regain their coherence though the relativistic corrections
destroy the coherence of harmonic potential in the
non-relativistic Schr\"{o}dinger equation.
 \item[ii.] For both small and large eigenvalues the minimum
uncertainties do not increase unbounded with time.  In fact this
means that wave packet does not spread unbounded.
 \item[iii.] The expectation
values of $x$ and $p$ for large $\alpha$ oscillate with
decreasing amplitude and approach zero which is equal to the
values of $\langle x\rangle$ and $\langle p\rangle$ in the energy
basis.
\end{itemize}
 We also introduced the relativistic Poschl-Teller potential and showed that
 its energy levels are equally-spaced, Eqs.(\ref{pt1},\ref{pten}). We obtained the
eigenfunctions of the annihilation operator for this potential,
Eq.(\ref{alpt}). It is consequently shown that the time evolution
of the obtained functions are still eigenstates of the
annihilation operator, Eq.(\ref{alptt}).  \\
 {\bf Acknowledgment}
The authors would like to thank B. Ferdowsian for his
contribution to this work at preliminary stage. Financial support
of Isfahan University of Technology and Institute for Studies in
Theoretical Physics and Mathematics (IPM) is gratefully
acknowledged.


\begin{thebibliography}{99}

\bibitem{glaub} R. J. Glauber, Phys. Rev. 131 (1963) 2766.
\bibitem{nieto} M. Martin Nieto, Phys. Rev. A17 (1978) 1273, Phys. Rev. A20 (1979) 700.
M. Martin Nieto and D. Rodney Truax, New J. Phys. 2 (2000) 18.1.
\bibitem{niesi} M. Martin Nieto and L. M. Simmons, Phys. Rev. D20 (1979) 1332.
\bibitem{gut} V. P. Gutschick and M. Martin Nieto, Phys. Rev. D22 (1980) 403.
\bibitem{chr} C. G. Christopher, J. Phys. A17 (1984) 737.
\bibitem{fuk}T. Fukui and N. Aizawa, Phys. lett. A180 (1978) 308.
\bibitem{zha} W. M. Zhang, D. H. Feng and R. Gilmore, Rev. Mod. Phys. 62 (1990) 867.
\bibitem{coop}I. L. Cooper, J. Phys. A25 (1992) 1671.
\bibitem{maj} P. Majumdar and H. S. Sharatchandra, quant-ph/9708010.
\bibitem{mol} B. Molnar and M. G. Benedict, quant-ph/9812041.
\bibitem{nou} S. Nouri, Phys. Rev. A60 (1999) 1702.
\bibitem{mat} M. Mathur and D. Sen, J. Math. Phys. 42 (2001) 4181.
\bibitem{ald} V. Aldaya and Gucrrcro, J. Math. Phys. 36 (1995) 3191.
\bibitem{klu} B. I. Lev, A. A. Semenov, C. V. Usenko and J. R.
Klauder, Phys. Rev. A66 (2002) 022115.
\bibitem{Doni} R. Dornhaus, G. Nimtz and B. Schlicht, Narrow-Gap
Semiconductors, Springer Tracts in Modern Physics Vol.98,
(Springer-verlag, 1983).  G. Junker, Supersymmetric Methods in
Quantum and Statistical Physics, (Springer-verlag, 1997).
\bibitem{dom}F. Dominguez-Adame, Phys. Lett. A136 (1989) 175.
\bibitem{grein} W. Greiner, Relativistic Quantum Mechanics (Springer-verlag, 1990).
\bibitem{pos} G. Poschl and E. Teller, Z. Phys. 83 (1933) 143.
\bibitem{sudar} V. I. Man'ko, G. Marmo, E. C. G. Sudarshan and F.
Zaccaria, Phys. Scr. 55 (1997) 528.
\end{thebibliography}
\end{document}